\documentclass[9pt]{article-hermes}
\usepackage{graphicx}
\usepackage{amsmath, amsthm, amssymb}
\usepackage{subfigure}

\title[Un Algorithme de Gestion des Adjacences basé sur la Puissance du Signal] {Un Algorithme de Gestion des Adjacences \\basé sur la Puissance du Signal\footnote{This work is supported by the French government funded project ANR RNRT R2M (Reseaux Mesh et Mobilite)}}

\author{Husnain Mansoor Ali        \andauthor Anthony Busson    \andauthor Amina Meraihi Naimi  \andauthor Veronique Veque}

\address{
University Paris XI\\
IEF - CNRS UMR 8622\\
Centre Scientifique d'Orsay\\
91405 Orsay - France\\[3pt]
\{husnain.ali,anthony.busson,amina.meraihi,veronique.veque\}@ief.u-psud.fr\\
}

\resume{Dans cette proposition, nous présentons une technique de gestion de lien pour les protocoles proactifs de routage dans les réseaux ad-hoc. Ce nouveau mécanisme de type multicouche est basé sur la puissance du signal. Le mécanisme d'hystérésis prévu par le protocole OLSR est amélioré par l'utilisation de la puissance du signal combinée avec les pertes de messages Hello. La puissance du signal est utilisée pour déterminer si la qualite de lien s'améliore ou est en baisse. Cela nous aide, non seulement à rendre la gestion de liens plus robuste, mais également à anticiper la rupture du lien et donc à améliorer les performances.}
\abstract{In this proposition, we present a link management technique for pro-active routing protocols for ad-hoc networks. This new mechanism is based on signal strength hence cross layer approach is used. The hysteresis mechanism provided by OLSR is improved upon by using signal strength in combination with the hello loss based hysteresis. The signal power is used to determine if the link-quality is improving or deteriorating while packet losses are handled through the hysteresis mechanism specified in OLSR RFC. This not only makes the link management more robust but also helps in anticipating link breakages thereby greatly improving the performance.}

\motscles{Protocoles de Routage, Gestion de Lien, OLSR, MANET, Puissance du Signal}

\begin{document}
\submitted{CFIP 2008}{1}
\proceedings{CFIP 2008}{0}
\maketitle

\section{Introduction}

	Dans les réseaux sans-fil, la mobilité des noeuds cause la rupture des liens radios. Cela peut avoir pour conséquences une dégradation importante des performances surtout en présence de mobilité rapide. 
Dans les réseaux ad-hoc, la gestion des adjacences, c'est-à-dire l'établissement de la liste des autres noeuds à portée radio est réalisé localement par chaque noeud.
La gestion de ces adjacences est effectué par le protocole de routage. Cette tâche est généralement effectué  sur la base de la réception ou non réception de messages de contrôles (message \texttt{Hello}). 
Donc habituellement, la puissance du signal pour la réception des messages \texttt{Hello} n'est pas considéré, en effet la couche réseau fait très peu d'hypothèse quant à la technologie sous-jacente.
Mais cela peut entraîner des taux de perte important~\cite{raisinghani-crosslayer}, lorsque le wifi est utilisé notemment.  
Aussi il existe quelques approches où la puissance du signal est prise en compte, mais cette approche est utilisé principalement dans le contexte des protocoles de routage réactifs~\cite{Crisostomo2004}. 
Dans cet article, nous nous focalisons plutôt sur les protocoles de routage proactifs. Avec un protocole proactif, les tables de routage contiennent une entrée pour chaque destination dans le réseau. Il ne nécessite donc pas de requêtes pour déterminer la route lors d'une nouvelle communication.

Dans cet article, nous proposons un algorithme efficace de gestion des adjacences dans le cadre d'une mobilité rapide (au delà de 60km/h) prenant en compte la puissance du signal. On suppose de plus qu'il s'agit d'un réseau mesh. 
Dans ce contexte, un réseau mesh est un réseau ad hoc où un certain nombre de noeuds, en principe statique, ont été déployés pour effectuer le relayage, ces noeuds accroissent la connectivité du réseau ad hoc et offrent plusieurs chemins possibles entre les sources et les destinations. 

Notre algorithme est une modification de l'algorithme de gestion des adjacences d'OLSR (Optimized Link State Routing Protocol)~\cite{olsr}. Cependant, il peut être utilisé par n'importe quel protocole de routage. 
Le problème avec l'algorithme initial est qu'il ne prend en compte que le fait que les paquets soient reçus ou perdus. Pour éviter de dévalider un lien à la moindre perte de message \texttt{Hello}, qui peut être provoqué par des collisions ou un évanouissement radio important, il y a un mécanisme d'hystérésis qui fait qu'il faut 2 pertes de \texttt{Hello} succéssives (avec les paramètres par défaut) pour considérer un lien invalide.
Cet hystérésis permet de stabiliser les liens radio, et est donc nécessaire pour palier aux pertes dûes à des phénomènes transitoires (collisions, evanouissement de la puissance radio). 
Mais lorsque deux noeuds s'éloignent l'un de l'autre, il faudra environ 4 secondes avant de considérer le lien invalide. Ces 4 secondes correspondent au temps entre le moment où les noeuds ne sont plus à portée radio et l'émission de deux \texttt{Hello} (qui ne sont pas reçus). 
Durant cette période, les deux noeuds considèrent le lien valide et peuvent émettre des données sur celui-ci. Toutes les données émises seront alors perdues. 
L'amélioration que nous apportons au mécanisme natif d'OLSR est la prise en compte de la détéroriation ou amélioration du signal. Nous n'attendons pas la perte sèche d'un ou deux \texttt{Hello} pour dévalider un lien, mais nous le dévalidons si durant plusieurs \texttt{Hello} succéssifs il y a une perte signicative de la puissance du signal. 

Dans la Section suivante, nous présentons notre algorithme, puis Section~\ref{sec:resultats} nous donnons quelques résultats de simulations. Nous concluons Section~\ref{sec:conclusion}.

\section{Algorithme de gestion des adjacences}

Comme dans OLSR, on associe à chaque lien une métrique de qualité. Cette métrique de qualité est sans unité.  
Lorsque cette métrique passe en dessous d'un seuil bas le lien n'est plus utilisé, lorsqu'elle passe au dessus d'un seuil haut le lien devient de nouveau utilisable. 
Dans OLSR, cette métrique est mis à jour uniquement en fonction des messages \texttt{Hello} perdus ou reçus. 
L'idée de notre algorithme est assez simple. Il s'agit d'augmenter la qualité du lien lorsque la qualité du signal radio augmente et inversement.
Sous l'hypothèse que la puissance du signal décroit avec la distance, on augmente la métrique du lien lorsqu'un noeud se rapproche et on la diminue quand un noeud s'éloigne de manière à dévalider le lien juste avant que le noeud sorte de la portée radio.  On utilise alors deux seuils (\textit{ss\_threshold\_low} et \textit{ss\_threshold\_high}) associés à la puissance du signal (et non pas à la métrique du lien).  
Lorsque la puissance du signal du \texttt{Hello} reçu est en dessous du seuil bas (\textit{ss\_threshold\_low}), on diminue la métrique de qualité. 
Lorsque la puissance du signal est entre les deux seuils, on est à la limite de la portée radio, et on prend en compte la puissance du signal pour savoir si le noeud s'éloigne ou se rapproche. 
Si la puissance du signal s'est nettement améliorée on augmente la qualité du lien, si au contraire il y a une détérioration nette on diminue la qualité du lien. 
Enfin, lorsque la puissance du signal est au dessus du seuil haut (\textit{ss\_threshold\_high}), on a une bonne qualité de lien radio et on augmente la métrique du lien (qu'il y ait dégradation ou non).

Nous donnons ci-dessous, l'algorithme en détail.

\newtheorem*{algorithm}{Signal Strength(ss) based Hysteresis Algorithm}
\begin{algorithm}

. \\IF there does not exist an entry in neighbor table \\
\hspace*{0.45cm}	Initialize Link\_quality \\
\hspace*{0.45cm}	Link\_pending $\gets$ true \\
\hspace*{0.45cm}	Sum\_sig\_var = 0 ; (to accumulate the signal variation) \\
ELSE \quad (if there is already an entry) \\
\hspace*{0.45cm}	IF ss $>$ ss\_threshold\_high ; (good reception; we reward) \\
\hspace*{0.45cm} \hspace*{0.45cm}	Link\_quality $\gets$ (1 - Hyst\_ss\_scaling ) * Link\_quality + Hyst\_ss\_scaling  \\
\hspace*{0.45cm}	ELSE (punish, reward or do nothing based on ss)\\
\hspace*{0.45cm} \hspace*{0.45cm}   IF Link\_pending $ = $ false AND (Sum\_sig\_var $ += Last\_ss-ss) \geq \Delta $ \\
\hspace*{0.45cm} \hspace*{0.45cm} \hspace*{0.45cm}(punish; signal strength deterioration $\geq  \Delta $)\\
\hspace*{0.45cm} \hspace*{0.45cm} \hspace*{0.45cm}	Link\_quality $ = Hyst\_ss\_scaling * Link\_quality $ \\
\hspace*{0.45cm} \hspace*{0.45cm} \hspace*{0.45cm}	Sum\_sig\_var $ = $ 0\\
\hspace*{0.45cm} \hspace*{0.45cm}	ENDIF \\
\hspace*{0.45cm} \hspace*{0.45cm}	IF Link\_pending $ = $ true AND (Sum\_sig\_var $ += ss-Last\_ss) \geq \Delta $ \\
\hspace*{0.45cm} \hspace*{0.45cm} \hspace*{0.45cm}	(reward; signal strength improvement $\geq \Delta $) \\
\hspace*{0.45cm} \hspace*{0.45cm} \hspace*{0.45cm}	Link\_quality $ = $ min(HYST\_THRESHOLD\_HIGH , $ (1-Hyst\_ss\_scaling) * Link\_quality + Hyst\_ss\_scaling $ ) \\
\hspace*{0.45cm} \hspace*{0.45cm} \hspace*{0.45cm}	Sum\_sig\_var $ = $ 0 \\
\hspace*{0.45cm} \hspace*{0.45cm}	ENDIF \\
\hspace*{0.45cm}	ENDIF \\
\hspace*{0.45cm}	Change the status of Link if Link\_quality crosses any of the two thresholds and reinitialize Sum\_sig\_var\\
ENDIF
\end{algorithm}

Dans l'algorithme ci-dessus, \textit{ss} est la valeur de la puissance du signal du dernier \texttt{Hello} reçu; \textit{Last\_ss} est la puissance de l'avant dernier \texttt{Hello} reçu et \textit{Hyst\_ss\_scaling} est un paramètre permettant de jouer sur la façon dont on augmente ou diminue la métrique de qualité du lien.
Enfin $\Delta$ est le paramètre auquel est comparé les différences de puissance. Plus précisement, l'algorithme augmente ou diminue la qualité du lien si il y a une améloriation/détérioration supérieure à $\Delta$ de la puissance du signal.
Cette amélioration/déterioration peut courir sur plusieurs \texttt{Hello} succéssifs.

\subsection{Résultats de simulations}\label{sec:resultats}
\label{simulation}

Pour nos simulations, nous utilisons une implémentation de OLSR du simulateur réseau NS version 2~\cite{ns}.
Nous montrons ici uniquement les résultats pour une topologie en chaîne.  
Elle consiste en une série de 10 noeuds statiques placés à 100 mètres les uns des autres. 
Un noeud mobile est initialement placé à 10 mètres du premier noeud statique, et se met en mouvement de manière à passer devant toute la chaîne de noeuds. 
La vitesse du noeud mobile varie de 20km/h à 100km/h. 
Le noeud mobile transmet alors deux paquets de 512 octets toutes les secondes au premier noeud de la chaîne. 
Nous comparons le PDR (Packet Delivery Ratio) définit comme le nombre de paquets reçus divisé par le nombre de paquets émis, pour l'algorithme d'OLSR (basé sur l'hystérésis) et notre propre algorithme.
Chaque point est la moyenne de $8$ simulations. A ces moyennes nous associons un intervalle de confiance à 95\% (une présentation plus exhaustives des résultats est donnée dans~\cite{mypaper}). 


\begin{figure}[tbp]
  \centering
    \leavevmode
    \subfigure[Comparaison du PDR]{
\fbox{\includegraphics[angle=-90,width=.45\linewidth]{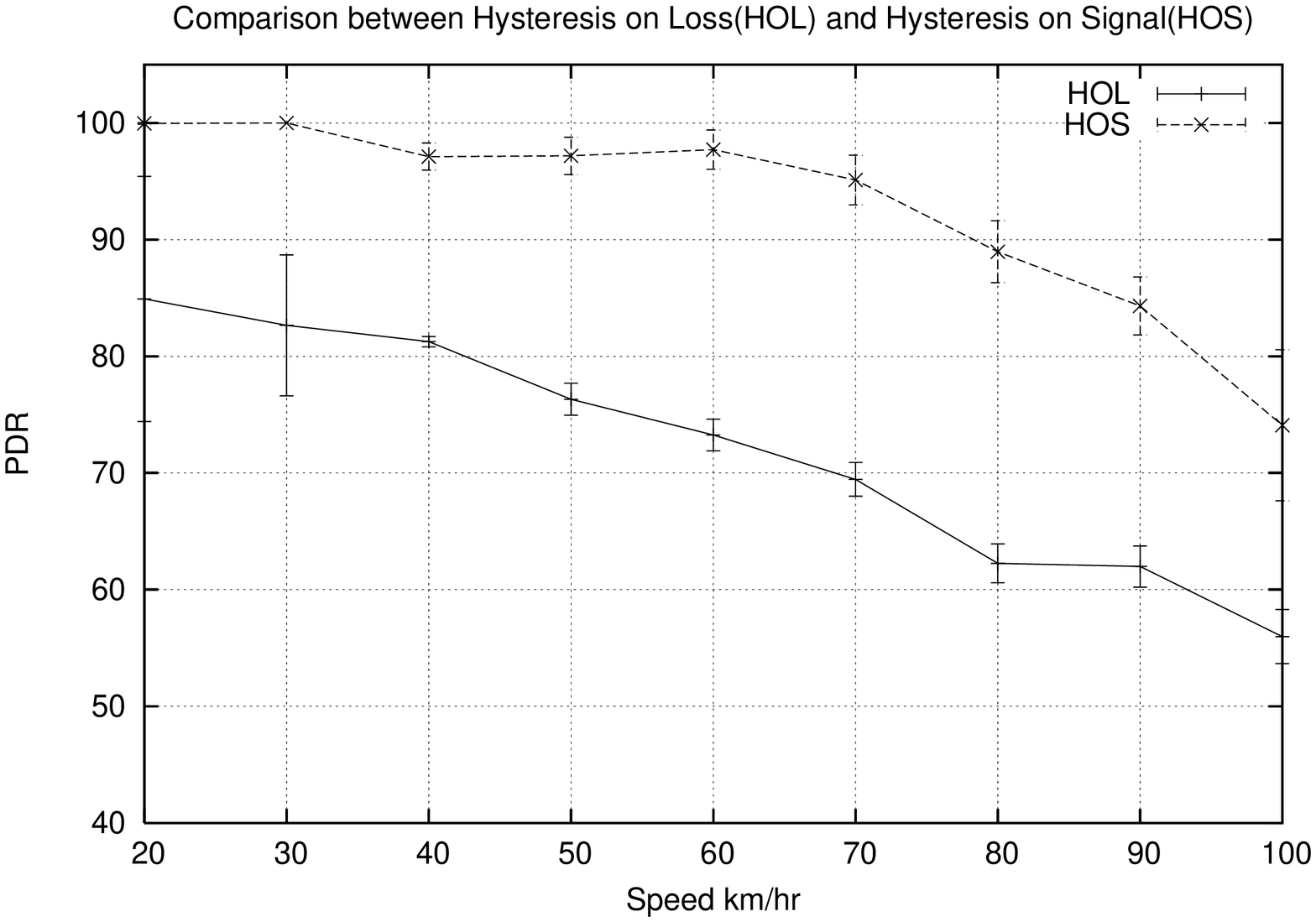}}
        \label{fig:chain}}
   \subfigure[Comparaison de l'overhead]{
\fbox{\includegraphics[angle=-90,width=.45\linewidth]{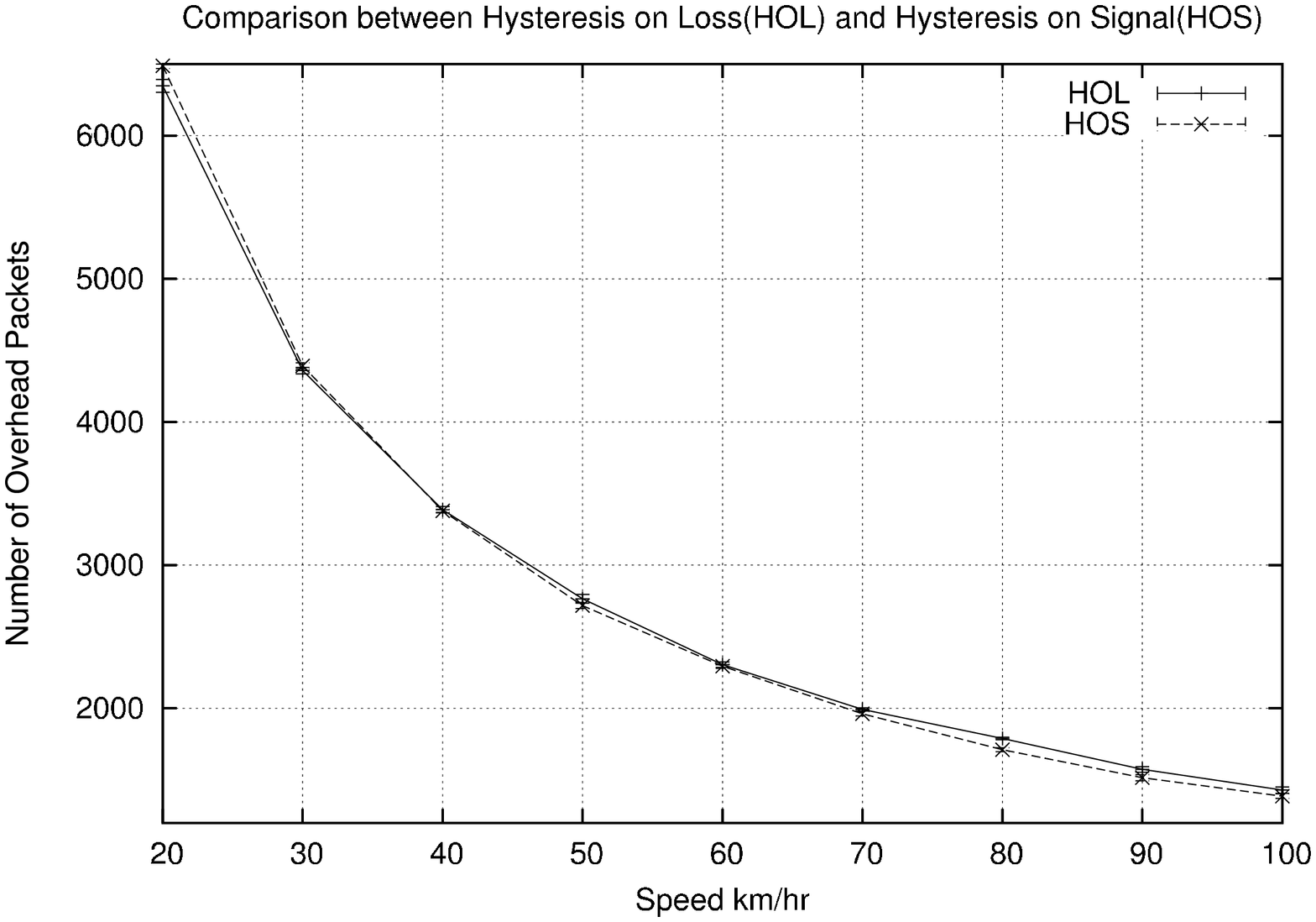}}
        \label{fig:overhead}}
   \caption{Comparaisons des PDR et des overhead pour le protocole natif d'OLSR (\textit{Hysteresis on Loss}) et notre algorithme (\textit{Hysteresis on Signal}).}
\end{figure}

Les résultats de simulations sont présentés sur les figures~\ref{fig:chain} et~\ref{fig:overhead}.
Notre algorithme est clairement plus performant que celui d'OLSR (noté \textit{Hysteresis on Loss} sur les figures).
Comme expliqué plus haut, la raison est qu'un noeud statique doit attendre 4 secondes après que le noeud mobile soit sortie de sa portée radio pour que le lien soit dévalider. 
On voit que pour notre algorithme, il y a un PDR proche de 100\% excepté pour les grandes vitesses. Ceci est dû au fait que notre algorithme a été dimensionné pour des vitesses moyennes (60km/h). 
La comparaison de l'overhead montre que notre algorithme n'utilise pas plus de paquets de contrôle que le OLSR du RFC. 
En effet, nous n'utilisons pas de nouveau message de contrôle et n'augmentons pas la fréquence des \texttt{Hello}.

\section{Conclusion}\label{sec:conclusion}

	Les résultats de simulations ont montré que la prise en compte de la puissance du signal rendait la gestion des adjacences plus robuste et permettait d'anticiper les ruptures de liens, diminuant ainsi nettement les pertes de paquets. 
De plus, celui-ci ne produit pas de messages de contrôle supplémentaire. Les travaux en cours ont pour but de donner des règles de dimensionnement de notre algorithme. En effet, les performances sont étroitement liées aux différents paramètres de l'algorithme, à la distance entre les noeuds statiques et à la vitesse des noeuds mobiles. 
Une comparaison de notre algorithme et des autres protocoles de gestion des adjacences prenant en compte la puissance du signal est également planifié.

\end{document}